\documentclass[10pt,twocolumn,conference]{IEEEtran}
\usepackage{setspace}
\usepackage{clrscode}
\usepackage{pict2e}

\usepackage{amsmath}
\usepackage{amssymb}
\usepackage[dvips]{graphicx}
\usepackage{epsfig}
\usepackage{algorithm}
\usepackage{algorithmic}
\usepackage{caption}
\usepackage{amsthm}
\usepackage{subcaption}

\usepackage[ps2pdf,
bookmarks=false,
bookmarksnumbered=false, 
bookmarksopen=false, 
colorlinks=false]{}

\makeatletter

\newcommand{\Rmnum}[1]{\expandafter\@slowromancap\romannumeral #1@}
\makeatother

\newcommand{\BS}{\text{BS}_0}
\newcommand{\BSR}{\text{BS}_0^R}
\newcommand{\RS}{\text{RS}_0}

\newcommand{\MUnc}{\text{MU}_{nc}}
\newcommand{\MUc}{\text{MU}_{c}}
\newcommand{\SINR}{\text{SINR}}

\newcommand{\g}{\text{g}}

\newcommand{\PC}{\text{P}_{\text{C}}}

\newcommand{\BE}{\text{BE}}

\begin{document}
%
\title{Energy Efficiency in Relay-Assisted mmWave Cellular Networks}

\author{Esma Turgut and M. Cenk Gursoy
\\
Department of Electrical Engineering and Computer Science\\
Syracuse University, Syracuse, NY 13244\\ Email:
eturgut@syr.edu, mcgursoy@syr.edu}

\maketitle

\begin{abstract}
In this paper, energy efficiency of relay-assisted millimeter wave (mmWave) cellular networks with Poisson Point Process (PPP) distributed base stations (BSs) and relay stations (RSs) is analyzed using tools from stochastic geometry. The distinguishing features of mmWave communications such as directional beamforming and having different path loss laws for line-of-sight (LOS) and non-line-of-sight (NLOS) links are incorporated into the energy efficiency analysis. Following the description of the system model for mmWave cellular networks, coverage probabilities are computed for each link. Subsequently,  average power consumption of BSs and RSs are modeled and energy efficiency is determined in terms of system parameters. Energy efficiency in the presence of beamforming alignment errors is also investigated to get insight on the performance in practical scenarios. Finally, the impact of BS and RS densities, antenna gains, main lobe beam widths, LOS interference range, and alignment errors on the energy efficiency is analyzed via numerical results.
\end{abstract}

\begin{IEEEkeywords}
Energy efficiency, millimeter wave cellular networks, Poison point processes, relay stations, stochastic geometry.
\end{IEEEkeywords}

\thispagestyle{empty}


\section{Introduction}
Recent years have witnessed exponential growth in mobile data and traffic due to, e.g., ever increasing use of smart phones, portable devices, and data-hungry multimedia applications. Limited available spectrum in microwave ($\mu$Wave) bands does not seem to be capable of meeting this demand in the near future, motivating the move to new frequency bands. Therefore, the use of large-bandwidth at millimeter wave (mmWave) frequency bands, between 30 and 300 GHz, becomes a good candidate for the fifth generation (5G) cellular networks and has attracted considerable attention recently \cite{Rappaport1} -- \cite{Ghosh}.

Despite the great potential of mmWave bands, increase in free-space path loss with increasing frequency, and poor penetration through solid materials such as concrete and brick, have made mmWave frequencies to be considered attractive only for short-range indoor communication. However, recent channel measurements have shown that these high frequencies may also be used for outdoor communication over a transmission range of about 150-200 meters \cite{Rappaport1}. It has also been shown that mmWave networks can achieve comparable coverage area and much higher data rates than $\mu$Wave networks when the base station density (BS) is sufficiently high and highly directional antennas are used \cite{Bai2}. With increase in the number of BSs in mmWave networks, however, energy efficiency is becoming an important consideration as well.

Energy efficiency of cellular networks has been extensively studied recently. Use of relay stations (RS) has been considered an effective way to have energy efficient and flexible networks while maintaining the coverage area and date rates. Unlike the BSs, RSs are not connected to the core network with wired backhaul, and therefore this provides a significant reduction in energy consumption. In \cite{Yu}, energy efficiency of relay-assisted networks are investigated using stochastic geometry. Authors of \cite{Wei} analyzed the effect of station density on the energy efficiency of relay-assisted cellular networks. However, these studies cannot be directly applied to mmWave cellular networks since unique features of mmWave communication have not been considered. Energy efficiency of millimeter wave cellular networks is studied in \cite{Rappaport2} and \cite{Yun}. In \cite{Rappaport2}, the impact of mmWave cellular channels on data rates and power consumption is analyzed using consumption factor framework. In \cite{Yun}, employment of RSs are combined with mmWave channel model. However, these two papers have not taken into account, in their energy efficiency analysis, the network model based on stochastic geometry. Therefore, in this paper, we employ stochastic geometry to analyze the energy efficiency of relay-assisted downlink mmWave cellular networks.

The rest of the paper is organized as follows. System model is introduced in Section II. In Section III, signal-to-interference-plus-noise ratio (SINR) coverage probabilities are computed for each link. In Section IV, energy efficiency is formulated initially considering perfect beam alignment, and then in the presence of beamsteering errors. In Section V, numerical results are presented to identify the impact of several system parameters on the energy efficiency. Finally, conclusions and suggestions for future work are provided in Section VI.

\section{System Model} \label{sec:system_model}
In this section, we introduce our system model for the relay-assisted downlink mmWave cellular network. The locations of BSs and RSs are modeled according to two independent homogeneous Poison Point Processes (PPPs) $\Phi_{B}$ and $\Phi_R$ of densities $\lambda_{B}$ and $\lambda_R$, respectively, on the Euclidean plane. Mobile users (MUs) are distributed according to some independent stationary point process. Two different types of MUs are considered in this paper: non-cooperative MU ($\MUnc$) and cooperative MU ($\MUc$). $\MUnc$s directly communicate with the serving BS which we denote by $\BS$, while $\MUc$s communicate with the serving BS via the help of the RSs. It assumed that the MUs are served by the closest nodes in the network. Let $\BS$ and $\RS$ be the closest base station and the closest relay, respectively, to a typical MU. MU is classified as $\MUnc$ if its distance to $\BS$ is less than that to $\RS$. Similarly, it is designated as a $\MUc$ if $\RS$ is closer to this MU than $\BS$. Also, RSs are associated with the closest BS, denoted by $\BSR$.

As shown in Fig. 1, $\BS$-$\MUnc$ and $\BS$-$\RS$-$\MUc$ links work in non-overlapping frequency bands with bandwidths $B_{nc}$ and $B_c$, respectively. A two-slot synchronous communication protocol is assumed in each cell for the $\BS$-$\RS$-$\MUc$ link. In the first time slot, BSs transmit signals to RSs, while in the second time slot, RSs forward the data (decoded from the received signal in the first time slot) to the $\MUc$s. The time duration of both time slots are assumed to be equal. Since separate frequency bands are assumed, the other-cell interference at $\MUnc$ is due to the BSs that use the same resource block with $\BS$. Similarly, the other-cell interference at RSs is from the BSs operating at the same frequency with $\BSR$, and interference at $\MUc$ is due to the RSs using the same frequency with $\RS$.

\begin{figure}
  \centering
   \includegraphics[width=0.5\textwidth]{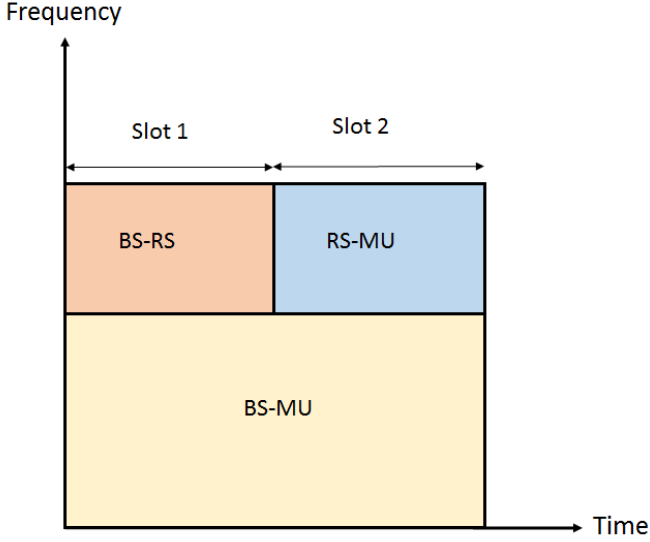}
\caption{Relay-assisted mmWave Cellular Network Frame Structure}
\label{fig:framestructure}
\end{figure}

In this setting, we have the following three assumptions regarding the system model of the downlink mmWave cellular network:

\textbf{Assumption 1:} Antenna arrays at the BSs, RSs and MUs are assumed to perform directional beamforming where the main lobe is directed towards the dominant propagation path while smaller sidelobes direct energy in other directions. For tractability in the analysis, antenna arrays are approximated by a sectored antenna model, in which the array gains are assumed to be constant $M$ for all angles in the main lobe and another smaller constant $m$ in the side lobe \cite{Hunter}. Initially, perfect beam alignment is assumed in the $\BS$-$\MUnc$, $\BSR$-$\RS$ and $\RS$-$\MUc$ links\footnote{Subsequently, beamsteering errors are also addressed.}, leading to an overall antenna gain of $MM$. Also, beam direction of the interfering nodes is modeled as a uniform random variable on $[0,2\pi]$. Therefore, the effective antenna gain is a discrete random variable (RV) described by

\begin{equation}
    G=\left\{
                \begin{array}{ll}
                  MM \quad \text{with prob.} \; p_{MM}=(\frac{\theta}{2\pi})^2\\
                  Mm \quad \text{with prob.} \; p_{Mm}=2\frac{\theta}{2\pi}\frac{2\pi-\theta}{2\pi}\\
                  mm \quad \text{with prob.} \; p_{mm}=(\frac{2\pi-\theta}{2\pi})^2,
                \end{array}
              \right. \label{eq:antennagains}
\end{equation}
where $\theta$ is the beam width of the main lobe, and $p_{G}$ is the probability of having an antenna gain of $G$.

\textbf{Assumption 2:} A BS or RS can either have a line-of-sight (LOS) or non-line-of-sight (NLOS) link to the MU according to the $LOS$ $probability$ $function$ $p(r)$ which is the probability that a link of length $r$ is LOS. Using field measurements and stochastic blockage models, $p(r)$ can be modeled as $e^{-\beta r}$ where decay rate $\beta$ depends on the building parameter and density \cite{Bai1}. For simplicity, $LOS$ $probability$ $function$ $p(r)$ can be approximated by a step function. In this approach, the irregular geometry of the LOS region is replaced with its equivalent LOS ball model with radius $R_B$ \cite{Bai2}. A BS or RS is a LOS node to the MU if it is inside the ball, otherwise it is a NLOS node. Different path loss laws are applied to LOS and NLOS links. Thus, the path-loss exponent on each interfering link can be expressed as follows:
\begin{equation}
    \alpha_i=\left\{
                \begin{array}{ll}
                  \alpha_L \quad \text{if} \; r \le R_B\\
                  \alpha_N \quad \text{if} \; r > R_B,
                \end{array}
              \right.  \label{eq:Pathloss}
\end{equation}
where $\alpha_L$ and $\alpha_N$ are the LOS and NLOS path-loss exponents, respectively.

\textbf{Assumption 3:} Serving nodes (BS or RS) are assumed to be LOS to the corresponding receiving nodes (RS or MU), and therefore the path loss exponent in the serving link is always equal to $\alpha_L$.

\section{Coverage Probability}
In this section, we first express the SINRs at the RSs and MUs by combining the above three assumptions with the described network model. Then, we derive the coverage probabilities for each link.

\subsection{Signal-to-Interference-plus-Noise Ratio (SINR)}
\textit{1)$\BS$-$\MUnc$ link:} The SINR in the downlink from the the $\BS$ to the $\MUnc$ can be written as:
\begin{equation}
\SINR_{BU}=\frac{P_{BU}G_0 h_0 x_0^{-\alpha_L}}{\sigma^2+\underbrace{\sum_{i \in \Phi_{B}^{\setminus{\BS}}}P_{BU}G_i h_i x_i^{-\alpha_i}}_{I_{BU}}}, \label{eq:SINR_BU}
\end{equation}
where $P_{BU}$ is the transmit power of $\BS$, $G_0$ is the effective antenna gain of the link which is assumed to be equal to $MM$, $h_0$ is the small-scale fading gain, $\alpha_L$ is the LOS path-loss exponent of the link, $x_0$ is the transmission distance, $\sigma^2$ is the variance of the additive white Gaussian noise component, and $I_{BU}$ is the aggregate other-cell interference at $\MUnc$. A similar notation is used for $I_{BU}$, but note that the effective antenna gain $G_i$ and path loss exponent $\alpha_i$ are different for different interfering links as described in (\ref{eq:antennagains}) and (\ref{eq:Pathloss}), respectively.

\textit{2)$\BSR$-$\RS$ and $\RS$-$\MUc$ links:} The SINRs in the downlink from the the $\BSR$ to the $\RS$, and from the $\RS$ to the $\MUc$ can be written, respectively, as follows:
\begin{equation}
\SINR_{BR}=\frac{P_{BR}G_0 g_0 y_0^{-\alpha_L}}{\sigma^2+\underbrace{\sum_{i \in \Phi_{B}^{\setminus{\BSR}}}P_{BR}G_i g_i y_i^{-\alpha_i}}_{I_{BR}}}, \label{eq:SINR_BR}
\end{equation}
\begin{equation}
\SINR_{RU}=\frac{P_{RU}G_0 \tilde{g}_0 \tilde{y}_0^{-\alpha_L}}{\sigma^2+\underbrace{\sum_{i \in \Phi_{R}^{\setminus{\RS}}}P_{RU}G_i \tilde{g}_i \tilde{y}_i^{-\alpha_i}}_{I_{RU}}}, \label{eq:SINR_RU}
\end{equation}
where a notation similarly as described in (\ref{eq:SINR_BU}) is used with similar parameter definitions.

All links are assumed to be subject to independent Nakagami fading (i.e., small-scale fading gains have a gamma distribution). Parameters of Nakagami fading are $N_L$ and $N_N$ for LOS and NLOS links, respectively, and they are assumed to be positive integers for simplicity.

\subsection{SINR Coverage Probability} \label{sec:SINR Coverage Probability}
The SINR coverage probability $\PC$ is defined as the probability that the received SINR is larger than a certain threshold $T>0$, i.e., $\PC= \mathbb{P}(\SINR>T)$. The coverage probability in the single-hop transmission and dual-hop relayed transmission can be formulated as follows:
\begin{equation}
\PC= \left\{
\begin{array}{ll}
\mathbb{P}(\SINR_{BU}>T) & \text{for}\;\MUnc \\
\mathbb{P}(\SINR_{BR}>T) \mathbb{P}(\SINR_{RU}>T) & \text{for}\;\MUc.
\end{array}
\right .
\end{equation}
Since decode-and-forward relaying strategy is employed by the RSs, a $\MUc$ is served if the SINRs of both links are larger than the threshold $T$. In other words, $\BS$-$\RS$-$\MUc$ link works if both RS and $\MUc$ can decode the received signal successfully.

Now, the coverage probability for the $\BS$-$\MUnc$ link can be calculated as
\begin{align}
\PC^{BU}&=\mathbb{P}(\SINR_{BU}>T) \nonumber \\
&=\int_{x_0>0} \mathbb{P}\bigg(h_0 \ge \frac{Tx_0^{\alpha_L}(\sigma^2+I_{BU})}{P_{BU}G_0} \mid x_0\bigg)f_{x_0}(x_0)dx_0 \nonumber \\
&= \int_{0}^{R_B} \sum_{n=1}^{N_L} (-1)^{n+1} {N_L \choose n} e^{-\frac{n \eta_L T x_0^{\alpha_L}\sigma^2}{P_{BU}G_0}} \times \nonumber \\
&\mathcal{L}_{I_{BU}}\bigg(\frac{n \eta_L T x_0^{\alpha_L}I_{BU}}{P_{BU}G_0} \bigg) f_{x_0}(x_0) dx_0, \label{CoverageofBUlink1}
\end{align}
where $f_{x_0}(x_0)=2\pi \lambda x_0 \exp\{-\pi\lambda x_0^2\}$ is the probability density function of the distance between an MU and its nearest LOS BS \cite{Andrews2}, $\eta_L=N_L(N_L!)^{-\frac{1}{N_L}}$, $\mathcal{L}_{I_{BU}}(s)$ is the Laplace transform of $I_{BU}$ evaluated at $s$, and (\ref{CoverageofBUlink1}) is derived noting that $|h_0|^2$ is a normalized gamma random variable with parameter $N_L$ and using the similar steps in \cite{Bai2}. Since $LOS$ $probability$ $function$ $p(\cdot)$ is equal to one inside the ball of radius $R_B$ and zero otherwise, integral in (\ref{CoverageofBUlink1}) is from 0 to $R_B$. We can employ the thinning property of PPP to split the $I_{BU}$ into 6 independent PPPs as
follows \cite{Bai3}:
\begin{align}
I_{BU} &= I_{BU,L} + I_{BU,N} \nonumber \\
&=I_{{BU,L}}^{MM}+I_{BU,L}^{Mm}+I_{BU,L}^{mm}+I_{BU,N}^{MM}+I_{BU,N}^{Mm}+I_{BU,N}^{mm} \nonumber \\
&=\sum_{G \in \{MM,Mm,mm\}}(I_{{BU,L}}^{G}+I_{{BU,N}}^{G}),  \label{eq:6PPP}
\end{align}
where $I_{BU,L}$ is the aggregate LOS interference arising from the BSs inside the LOS ball, $I_{BU,N}$ is the aggregate NLOS interference from outside the LOS ball, and $I_{{BU,L}}^{G}$ and $I_{{BU,N}}^{G}$ denote the LOS and NLOS interferences, respectively, with random antenna gain $G$ defined in (\ref{eq:antennagains}).
According to the thinning theorem, each independent PPP has a density of $\lambda_Bp_{G}$ where $p_{G}$ is given in (\ref{eq:antennagains}) for each
antenna gain $G \in \{MM, Mm, mm\}$.

Inserting (\ref{eq:6PPP}) into the Laplace transform expression and using the definition of Laplace transform yield
\begin{align}
\mathcal{L}_{I_{BU}}(s)= \mathbb{E}_{I_{BU}}[e^{-sI_{BU}}]=\mathbb{E}_{I_{BU}}[e^{-s(I_{BU,L}+I_{BU,N})}] \nonumber \\
&\hspace{-7.5cm}\stackrel{(a)}{=}\mathbb{E}_{I_{BU,L}}\big[e^{-s(I_{BU,L}^{MM}+I_{BU,L}^{Mm}+I_{BU,L}^{mm})}\big] \mathbb{E}_{I_{BU,N}}\big[{e^{-s(I_{BU,N}^{MM}+I_{BU,N}^{Mm}+I_{BU,N}^{mm})}}\big] \nonumber \\
&\hspace{-7.5cm}= \prod_G \prod_j \mathbb{E}_{I_{BU,j}^G}[ e^{-sI_{BU,j}^G}], \label{eq:LT}
\end{align}
where $G \in \{MM, Mm, mm\}$, $j \in \{L,N\}$, $s=\frac{n \eta_L T x_0^{\alpha_L}}{P_{BU}G_0}$, and (a) follows from the fact that  $I_{BU,L}$ and $I_{BU,N}$ are interferences generated from two independent thinned PPPs $\Phi_{B,L}$ and $\Phi_{B,N}$, respectively. Now, we can compute the Laplace transform for the LOS interfering links with a generic antenna gain $G$ using stochastic geometry as follows:
\begin{align}
\mathbb{E}_{I_{BU,L}^G}[ e^{-sI_{BU,L}^G}]&= e^{-2\pi\lambda_{B}p_{G} \int_{x_0}^{R_B}(1-\mathbb{E}_h [ e^{-s P_{BU}G h t^{-\alpha_L} }])p(t)t dt} \nonumber \\
&\stackrel{(a)}{=} e^{-2\pi\lambda_{B}p_{G} \int_{x_0}^{R_B} (1-1/(1+sP_{BU}Gt^{-\alpha_L}/N_L)^{N_L})t dt}, \label{eq:LT_LOS}
\end{align}
where $p(\cdot)$ is again the \emph{LOS probability function}, which is equal to 1 inside the ball and (a) is obtained by computing the moment generating function (MGF) of the gamma random variable $h$. Similarly, Laplace transform for the NLOS interfering links with a generic antenna gain $G$ can be calculated as
\begin{align}
\hspace{-.3cm}\mathbb{E}_{I_{BU,N}^G}[ e^{-sI_{BU,N}^G}]&= e^{-2\pi\lambda_{B}p_{G} \int_{R_B}^{\infty}(1-\mathbb{E}_h [ e^{-s P_{BU}G h t^{-\alpha_N} }])(1-p(t))t dt} \nonumber \\
&= e^{-2\pi\lambda_{B}p_{G} \int_{R_B}^{\infty} (1-1/(1+sP_{BU}Gt^{-\alpha_N}/N_N)^{N_N})t dt}. \label{eq:LT_NLOS}
\end{align}

By inserting (\ref{eq:LT_LOS}) and (\ref{eq:LT_NLOS}) into (\ref{eq:LT}), Laplace transform of $I_{BU}$ can be obtained. Finally, SINR coverage probability for the $\BS$-$\MUnc$ link is given at the top of the next page in (\ref{CoverageofBUlink2}).
\begin{figure*}
\scriptsize
\begin{equation}
\hspace{-1.2cm}\PC^{BU}=\!\int_{0}^{R_B}\!\! \sum_{n=1}^{N_L} (-1)^{n+1} {N_L \choose n} e^{-\frac{n \eta_L T x_0^{\alpha_L}\sigma^2}{P_{BU}G_0}} e^{-2\pi\lambda_{B} (\sum_{i=1}^3 p_{G_i} \int_{x_0}^{R_B} (1-1/(1+sP_{BU}Gt^{-\alpha_L}/N_L)^{N_L})t dt+ \sum_{i=1}^3 p_{G_i} \int_{R_B}^{\infty} (1-1/(1+sP_{BU}Gt^{-\alpha_N}/N_N)^{N_N})t dt)} e^{-\pi\lambda_B x_0^2}2\pi\lambda_B x_0 dx_0. \label{CoverageofBUlink2}
\end{equation}
\normalsize
\end{figure*}

SINR coverage probability for the $\BSR$-$\RS$ link can be computed by following similar steps, and it is given at the top of the next page in (\ref{CoverageofBRlink2})
\begin{figure*}
\scriptsize
\begin{equation}
\hspace{-1.3cm}\PC^{BR}=\!\int_{0}^{R_B}\!\! \sum_{n=1}^{N_L} (-1)^{n+1} {N_L \choose n} e^{-\frac{n \eta_L T y_0^{\alpha_L}\sigma^2}{P_{BR}G_0}} e^{-2\pi\lambda_{min} (\sum_{i=1}^3 p_{G_i} \int_{y_0}^{R_B} (1-1/(1+sP_{BR}Gt^{-\alpha_L}/N_L)^{N_L})t dt+ \sum_{i=1}^3 p_{G_i} \int_{R_B}^{\infty} (1-1/(1+sP_{BR}Gt^{-\alpha_N}/N_N)^{N_N})t dt)} e^{-\pi\lambda_B y_0^2}2\pi\lambda_B y_0 dy_0. \label{CoverageofBRlink2}
\end{equation}
\normalsize
\end{figure*}
where $s=\frac{n \eta_L T y_0^{\alpha_L}}{P_{BR}G_0}$, $\lambda_{min}=\min\{\lambda_B,\lambda_R\}$. The only difference is that in the derivation of the Laplace transform $\lambda_{min}$ is used instead of $\lambda_B$ because at any time only at most $\lambda_{min}$ BSs per square meter are transmitting signals to RSs.

For the $\RS-\MUc$ link, SINR coverage probability can be computed similarly as for the other links, but the distance between RS and MU follows a different distribution. Since the RSs are distributed according to a PPP and MUs follows some independent stationary point process in the given circular region around the RS, the distance between the MU and its corresponding RS follows a distribution with pdf $f_R(r)=2r/a^2$ for $0 \leq r \leq  a $ \cite{Yu}. Also, since only the RSs with received SINR larger than a certain threshold can decode and forward the signal to the $\MUc$s, the density used in SINR coverage calculation for this link is $\lambda^{\prime}=\lambda_{min}\PC^{BR}$. Finally, coverage probability for the $\RS-\MUc$ link is given at the top of the next page in (\ref{CoverageofRUlink}) where $s=\frac{n \eta_L T \tilde{y}_0^{\alpha_L}}{P_{RU}G_0}$.
\begin{figure*}
\scriptsize
\begin{equation}
\hspace{-1cm} \PC^{RU}=\int_{0}^{a} \sum_{n=1}^{N_L} (-1)^{n+1} {N_L \choose n} e^{-\frac{n \eta_L T \tilde{y}_0^{\alpha_L}\sigma^2}{P_{RU}G_0}} e^{-2\pi\lambda^{\prime} (\sum_{i=1}^3 p_{G_i} \int_{\tilde{y}_0}^{R_B} (1-1/(1+sP_{RU}Gt^{-\alpha_L}/N_L)^{N_L})t dt+ \sum_{i=1}^3 p_{G_i} \int_{R_B}^{\infty} (1-1/(1+sP_{RU}Gt^{-\alpha_N}/N_N)^{N_N})t dt)} \frac{2\tilde{y}_0}{a^2} d\tilde{y}_0. \label{CoverageofRUlink}
\end{equation}
\normalsize
\end{figure*}

\section{Energy Efficiency Analysis}
\subsection{Power Model}
The total power consumption per BS or RS can be modeled as $P_{tot}=P_0+\beta P_T$, where $1/\beta$ is the efficiency of the power amplifier, and $P_0$ is the static power consumption due to signal processing, battery backup, site cooling etc., and $P_T$ corresponds to the transmit power \cite{Richter}. Using this power formulation, the average power consumption of BSs (per unit area) in the cellular network can be expressed as
\begin{equation}
P_{B_{avg}}=\lambda_B P_{B_0}+ \beta_B(\lambda_B P_{BU}+\lambda_{min} P_{BR}/2), \label{eq:AvgPowers1}
\end{equation}
where  $P_{B_0}$ is the static power consumption of a BS, $1/\beta_B$ is the efficiency of power amplifiers at the BSs, and 1/2 factor is due to the fact that RSs are active only in one of the two time slots as depicted in Fig. \ref{fig:framestructure}. The first term is the average static power consumed regardless of whether the BSs are active or inactive, and the second term is the average transmit power consumed at BSs transmitting to $\MUnc$s and RSs. Note that at most only $\lambda_{min}=\min\{\lambda_B, \lambda_R\}$ BSs per square meter are transmitting signals to RSs.

Similarly, the average power consumption of the RSs (per unit area) in the cellular network is given by
\begin{equation}
P_{R_{avg}}=(\lambda_R-\lambda^{\prime})P_{R_0}+ \lambda^{\prime}(\beta_R P_{RU}/2+P_{R_0}),  \label{eq:AvgPowers2}
\end{equation}
where $P_{R_0}$ is the static power consumption of an RS, $1/\beta_R$ is the efficiency of power amplifiers at the RSs, and 1/2 factor is due to the fact that RSs are active only in the second time slot. In this scenario, only the RSs which can decode the signals from BSs can successfully forward them to $\MUc$s, and therefore the density of the active RSs is $\lambda^{\prime}=\lambda_{min}\PC^{BR}$. As a result, ($\lambda_R-\lambda^{\prime}$) RSs per square meter are inactive and they consume only static power. Thus, the second term is the sum of average transmit power and average static  power consumed at active RSs.

\subsection{Energy Efficiency Metric}  \label{sec:Energy Efficiency Function}
Energy efficiency can be measured and quantified as the ratio of the area spectral efficiency to the average network power consumption:
\begin{equation}
\text{EE}=\frac{\tau_{nc}+\tau_c}{P_{B_{avg}}+ P_{R_{avg}}} (bps/Hz/W)
\end{equation}
where $\tau_{nc}$ and $\tau_c$ are the area spectral efficiencies taken over all the BS-MU and BS-RS-MU links, respectively. The area spectral efficiency (i.e., network throughput) can be defined as the product of the throughput at a given link and density of transmitters (BSs or RSs), and can be formulated as follows \cite{Yu}:
\begin{align}
\tau_{nc}&= \frac{B_{nc}}{B_{nc}+B_c} \lambda_B \PC^{BU} \log_2(1+T) \\
\tau_{c}&= \frac{1}{2}\frac{B_{c}}{B_{nc}+B_c} \lambda_{min} \PC^{BR} \PC^{RU} \log_2(1+T), \label{eq:ASEofRS}
\end{align}
where $1/2$ factor is due to half-duplex operation of the RSs.

\subsection{Coverage and Energy Efficiency In the Presence of Beamsteering Errors}
In Section \ref{sec:Energy Efficiency Function} and the preceding analysis, antenna arrays at the serving nodes (BS or RS) and receiving nodes (RS or MU) are assumed to be aligned perfectly and energy efficiency is calculated in the absence of beamsteering errors. However, in practice, it may not be easy to have perfect alignment. Therefore, in this section, we investigate the effect of beamforming alignment errors on the energy efficiency of the network. We employ an error model similar to that in \cite{Wildman}. Let $|\epsilon|$ be the random absolute beamsteering error of the transmitting node toward the receiving node with zero-mean and bounded absolute error $|\epsilon|_{\text{max}} \le \pi$. Due to symmetry in the gain $G_0$, it is appropriate to consider the absolute beamsteering error. The PDF of the effective antenna gain $G_0$ with alignment error can be explicitly written as \cite{Marco2}
\begin{align}
f_{G_0}(\g)&=F_{|\epsilon|}\left(\frac{\theta}{2}\right)^2\delta(\g-MM)+2F_{|\epsilon|}\left(\frac{\theta}{2}\right)\!\!\left(\!\!1-F_{|\epsilon|}\left(\frac{\theta}{2}\right)\!\!\right) \times \nonumber \\
&\delta(\g-Mm) +\left(\!\!1-F_{|\epsilon|}\left(\frac{\theta}{2}\right)\!\!\right)^2\delta(\g-mm),
\label{eq:PDFofG}
\end{align}
where $\delta(\cdot)$ is the Kronecker's delta function, $F_{|\epsilon|}(x)$ is the CDF of the misalignment error and (\ref{eq:PDFofG}) follows from the definition of CDF, i.e., $F_{|\epsilon|}(x)=\mathbb{P}\{|\epsilon|\le x\}$. Assume that the error $\epsilon$ is Gaussian distributed, and therefore the absolute error $|\epsilon|$ follows a half normal distribution with $F_{|\epsilon|}(x)=\text{erf}(x/(\sqrt{2}\sigma_{\BE}))$, where $\text{erf}(\cdot)$ denotes the error function and $\sigma_{\BE}$ is the standard deviation of the Gaussian error $\epsilon$.

It is clear that all SINR coverage probability expressions in Section \ref{sec:SINR Coverage Probability} depend on the effective antenna gain $G_0$ between the serving and the receiving nodes, and so does the energy efficiency. Thus, SINR coverage probability $\PC$ for a generic link can be calculated by averaging over the distribution of $G_0$, $f_{G_0}(\g)$, as follows:
\begin{align}
\PC &= \int_0^{\infty}\PC(g)f_{G_0}(\g)d \g \nonumber \\
&=(F_{|\epsilon|}(\theta/2))^2 \PC(MM)+2(F_{|\epsilon|}(\theta/2))\bar{F}_{|\epsilon|}(\theta/2) \PC(Mm) \nonumber \\
&+\bar{F}_{|\epsilon|}(\theta/2)^2 \PC(mm),
\end{align}
where we define $\bar{F}_{|\epsilon|}(\theta/2)=1-F_{|\epsilon|}(\theta/2)$.

Applying this averaging to coverage probability expressions in all links, and inserting them to the area spectral efficiency and the average network power consumption formulas, we can obtain the modified energy efficiency expressions in the presence of beamsteering errors.

\section{Numerical Results}
In this section, numerical results are provided to analyze the impact of key system parameters on the energy efficiency of a downlink mmWave cellular network. We employ the parameter values listed in Table \ref{Table} unless stated otherwise.

\begin{table}
\small
\caption{System Parameters}
\centering
  \begin{tabular}{| l | r|}
    \hline
    Parameters & Values \\ \hline
    $\alpha_L$, $\alpha_N$& 2, 4 \\ \hline
    $N_L$, $N_N$& 3, 2 \\ \hline
    $M$, $m$ & 20dB, -10dB  \\ \hline
    $\lambda_R$ & $10^{-4}/m^2$ \\ \hline
    $R_B$, $a$ & 100m, 30m \\ \hline
    $T$, $\sigma^2$ & 30dB, -70dBm \\ \hline
    $B_{nc}$, $B_c$ & 1GHz, 100MHz  \\ \hline
    $P_{BU}$, $P_{BR}$, $P_{RU}$ & 50dBm, 50dBm, 30dBm  \\ \hline
    $P_{B_0}$, $P_{R_0}$ & 100W, 5W  \\ \hline
    $\beta_B$, $\beta_R$ & 5, 4 \\ \hline
  \end{tabular} \label{Table}
\end{table}

\begin{figure}
\centering
  \includegraphics[width=0.5\textwidth]{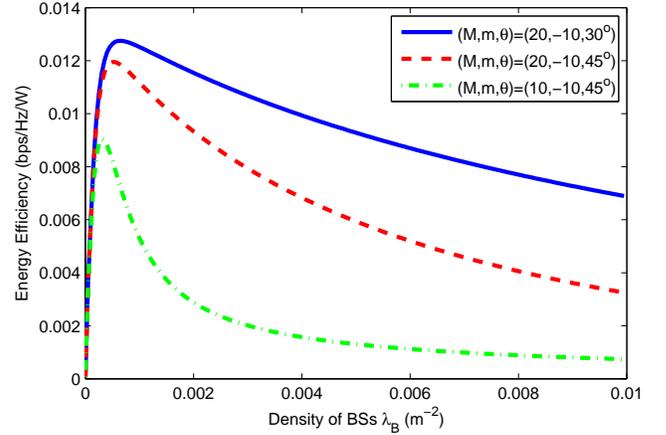}
  \caption{\small EE as a function of the BS density $\lambda_B$ for different antenna patterns ($M$,$m$,$\theta$). \normalsize}
\label{Fig1}
\end{figure}

First, we display the energy efficiency for different antenna patterns. We investigate the effect of the main lobe gain and the main lobe beam width in Fig. \ref{Fig1}. It can be seen that for fixed $\theta$, energy efficiency improves with increasing main lobe gain $M$. Similarly, for fixed $M$ decreasing $\theta$ improves energy efficiency because narrower main lobe beam width means that receiving nodes (RS or MU) are less likely to be interfered by the main lobe of other transmitting nodes (BS or RS). Also note that optimal BS density, denoted by $\lambda_{B}^{*}$, with which the energy efficiency is maximized, decreases slightly with increasing beam width and decreasing gain due to growing impact of interference.

\begin{figure}
\centering
  \includegraphics[width=0.5\textwidth]{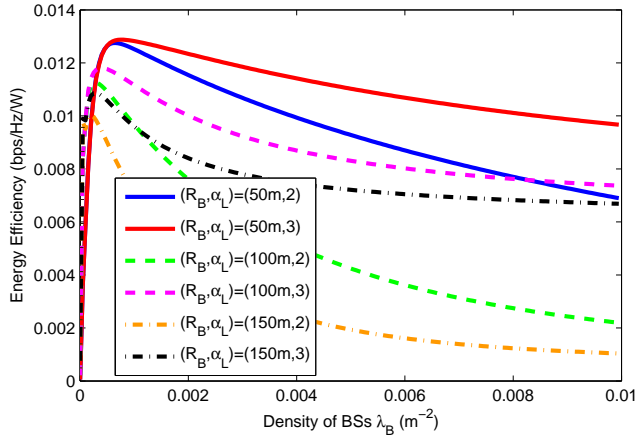}
  \caption{\small EE as a function of the BS density $\lambda_B$ for different LOS ball radiuses $R_B$ and LOS path-loss exponents $\alpha_L$. \normalsize}
\label{Fig3}
\end{figure}

Next, we plot the energy efficiency for different values of the LOS ball radius $R_B$ and LOS path loss exponent $\alpha_L$ in Fig. \ref{Fig3} in order to investigate the effect of LOS interference range. We notice that optimal BS density $\lambda_{B}^{*}$ decreases with increasing ball radius, because the number of interfering LOS BSs increases with increasing ball radius, and we have assumed that serving nodes are always LOS to the receiving nodes. As a result, the maximum energy efficiency is achieved with smaller BS density for higher ball radiuses. Also, energy efficiency improves with increasing LOS path loss exponent for fixed $R_B$, while $\lambda_{B}^{*}$ remains almost same for the same $R_B$. Therefore, optimal BS density is generally insensitive to the path loss exponent.

\begin{figure}
  \centering
  \includegraphics[width=0.5\textwidth]{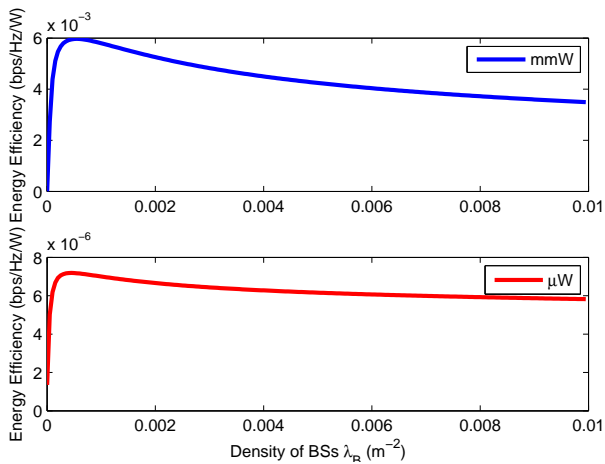}
  \caption{\small EE as a function of the BS density $\lambda_B$ for mmWave and $\mu$Wave ($\lambda_R=10^{-2}$, $T=20dB$).\normalsize}
 \label{Fig4}
\end{figure}

In Fig. \ref{Fig4}, we compare the performance of mmWave cellular network with that of microwave ($\mu$Wave) cellular network. For $\mu$Wave cellular networks, we set $M=m=0$dB, $R_B=1500$m, $\theta=2\pi$, $B_{nc}=50$MHz, $B_c=5$MHz. The rest of the parameters are the same as in the mmWave network setup. Top subfigure in Fig. \ref{Fig4} shows the performance of mmWave networks, while the bottom subfigure depicts the performance of the $\mu$Wave one. We immediately notice that mmWave networks have much higher energy efficiency than in $\mu$Wave networks (noting the scale differences in the y-axes of the two figures). This is mainly due to the use of omni-directional antennas and assuming all interfering nodes whose distances are less than $R_B=1500$m are interfering with the LOS path loss $\alpha_L  = 2$ in the $\mu$Wave cellular network setup, causing severe interference.

\begin{figure}
  \centering
  \includegraphics[width=0.5\textwidth]{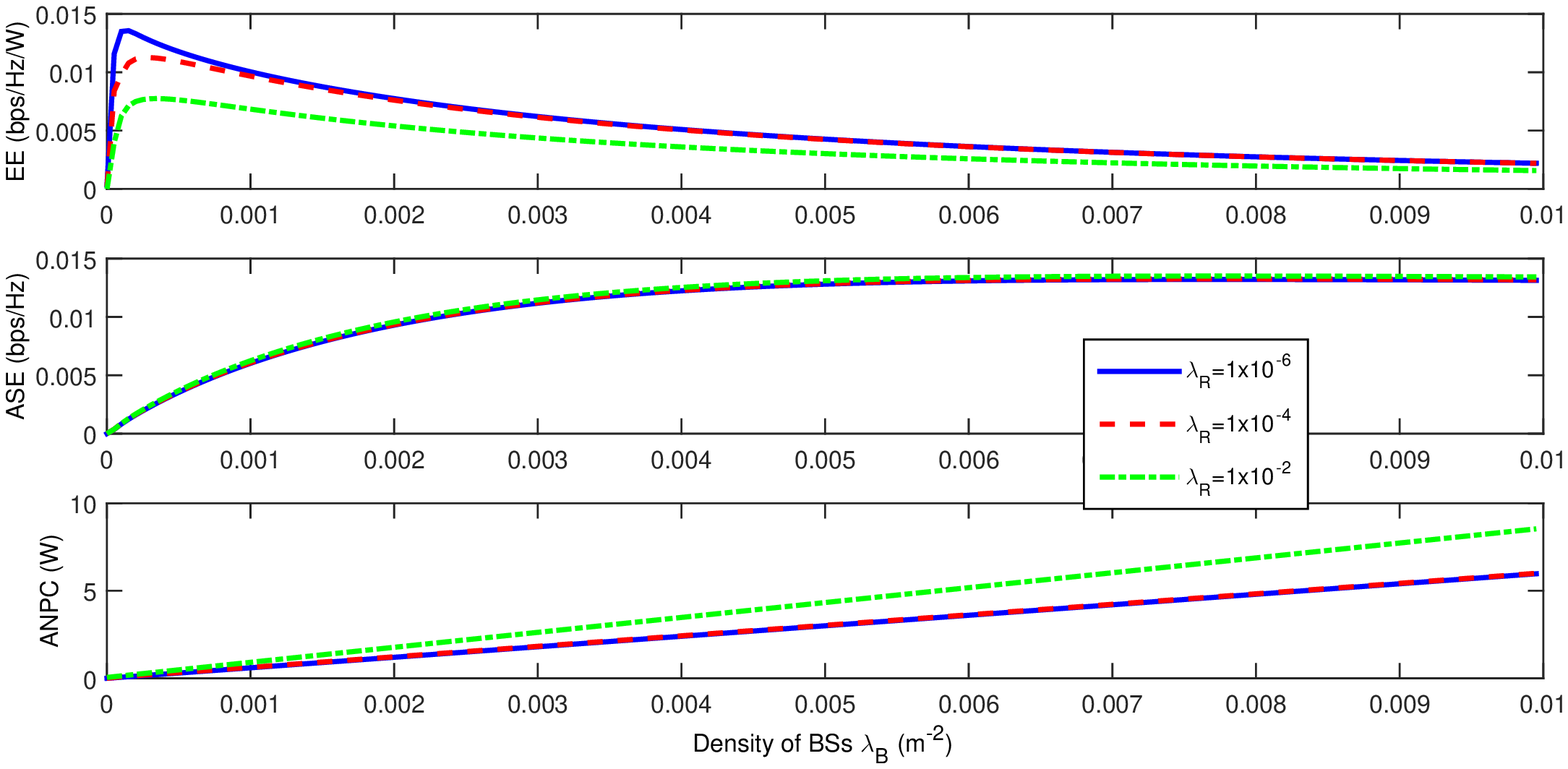}
  \caption{\small EE, ASE and ANPC as a function of the BS density $\lambda_B$ for different RS densities $\lambda_R$. \normalsize}
\label{Fig6}
\end{figure}

In Fig. \ref{Fig6}, we plot the energy efficiency (EE), area spectral efficiency (ASE) and average network power consumption (ANPC) as a function of the BS density $\lambda_B$ for different values of RS density $\lambda_R$, and investigate the effect of RS density on energy efficiency. As shown in the middle sub-figure, area spectral efficiency of the network increases only very slightly (which is difficult to notice in the figure but can be seen with higher resolution) with the increasing RS density because of the increase in SINR coverage probabilities. At the same time, however, having higher number of RSs means more power consumption. Consequently, average power consumption of the network also increases as shown in the bottom sub-figure. Since increase in area spectral efficiency cannot compensate for the increase in average network power consumption, energy efficiency degrades with increasing RS density as shown in the top sub-figure. This behavior indicates a tradeoff between area spectral efficiency and energy efficiency depending on the RS density.

\begin{figure}
  \centering
  \includegraphics[width=0.5\textwidth]{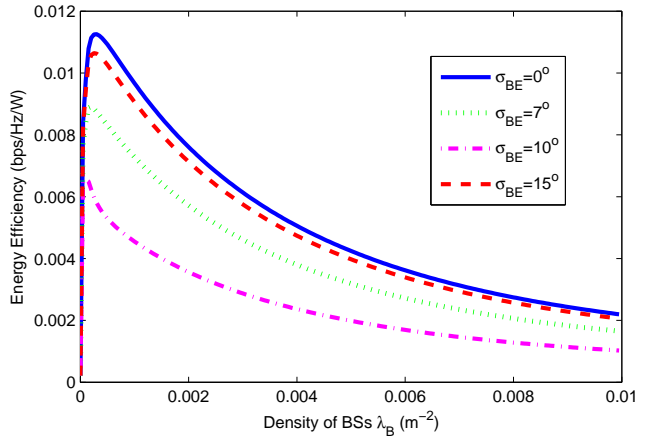}
  \caption{\small EE as a function of the BS density $\lambda_B$ for different alignment errors $\sigma_{BE}$. \normalsize}
 \label{Fig5}
\end{figure}

Finally, we investigate the effect of beam steering errors between the serving and receiving nodes on the energy efficiency in Fig. \ref{Fig5}. As shown in the figure, energy efficiency diminishes with increasing alignment error. Although the interference from interfering nodes remains unchanged, its effect grows with the increase in alignment error on the main link. Also, due to this increase in the relative impact of interference, less number of BSs is preferred with increasing alignment error to achieve the maximum energy efficiency.

\section{Conclusion}\label{sec:conclusion}
In this paper, we have analyzed the energy efficiency of relay-assisted downlink mmWave cellular networks by incorporating the distinguishing features of mmWave communication into the energy efficiency analysis. Directional beamforming with sectored antenna model and simplified ball-LOS models have been considered in the analysis. BSs and RSs are assumed to be distributed according to independent PPPs, and SINR coverage probabilities are derived using tools from stochastic geometry to characterize the energy efficiency. Numerical results demonstrate that employing directional antennas makes the mmWave cellular networks more energy efficient than $\mu$Wave cellular networks. In other words, increasing the main lobe gain and decreasing the main lobe beam width results in improved energy efficiency. We have also shown that BS density should be lowered to achieve the maximum energy efficiency when the LOS ball radius is larger. Moreover, we have observed that there is a tradeoff between the area spectral efficiency and energy efficiency depending on the RS density. Finally, the effect of alignment error on energy efficiency is quantified. Investigating the effect of using different LOS probability functions instead of the simplified ball-LOS model remains as future work.


\end{document}